# Soft Matter

# Stability, electronic, magnetic and thermoelectric properties of quaternary Heusler alloys CoX'ZrAl (X'=V, Fe, Ir): 3d Vs 5d systems[†]

Poulami Biswas,[a] Mahabubur Rahaman[b] and Molly De Raychaudhury[a,*]

The thermoelectric (TE) properties of quaternary Heusler alloys (CoX'ZrAl; X'=V, Fe, Ir) are studied in the framework of Density Functional Theory and Boltzmann Transport Theory. The compound CoVZrAl is found to be semiconducting and the most difficult to be formed whereas the easier formed newly-predicted CoFeZrAl and CoIrZrAl are pseudo-gapped and half-metal respectively. Ferromagnetic calculations show magnetism in CoVZrAl/CoIrZrAl originating from the non-bonding Co-X' fully-filled $t_{1u}$ state/partially filled $e_u$ state derived from Co-3d and X'-3d/5d electrons while CoFeZrAl is non-magnetic. The two-dimensional graphene-like density of states at the Fermi level in CoFeZrAl implies large electrical conductivity. We further observe that the presence of two 3d early Transition metal (TM) atoms enhances the Seebeck coefficient as in CoVZrAl and CoFeZrAl and the presence of extended 5d state of Ir diminishes the same for CoIrZrAl. However the higher thermal conductivity in CoFeZrAl renders CoVZrAl the best TE material among the three. The ZT value of n-type CoVZrAl is found to reach a higher value 1.4 at 600 K.

## 1 Introduction

Full Heusler alloys have been of much interest for a considerable length of time, especially for those interested in the search for spin injectors, spin filters and good thermoelectric materials. The first two terminologies, relevant for spintronics, need a ferromagnetic (FM) metal with maximum spin polarization to act as a spin injector or a ferromagnetic insulator to act as a spin filter. Several full Heusler alloys[1] are known to be good spin injectors. Transition-metal (TM) based full Heusler alloys are known for their half-metallicity, high Curie temperature ($T_c$) and scalable electronic properties[2]. Among these alloys, the most extensively investigated material is Co$_2$MnSi[3–21]. However a major disadvantage of these alloys is that the theoretical predictions of 100% spin polarization have often not been realized upon synthesis. This has been mainly due to structural or antisite disorder among the TM sites and spin disorder at finite temperatures[6,21–25].

So the search for full Heusler alloys that possess a 100% spin-polarization and that which can be realized in laboratory, continues till date. On the other hand, spin filter materials (SFM) are magnetic semiconductors which act as insulating barriers between magnetic electrodes. Through this insulating barrier, tunneling of electrons takes place, thereby leading to ballistic transport. Now the major setback here is that ferromagnetic semiconductors occur rarely in nature. Among the few known magnetic semiconductors are EuO, NiFe$_2$O$_4$, CoFe$_2$O$_4$, CoCrO$_4$ and BiMnO$_3$[26–30]. A new FM semiconductor CoVZrAl has already been predicted using first-principles calculation in recent times[31]. However this material could not be synthesized even a decade after its prediction for reasons unknown to us.

Another crucial area where these kinds of full Heusler alloys can find applicability is in the conversion of waste heat to electricity and vice versa, using their TE properties. Many Heusler alloys have been considered as potential TE material such as Co$_2$TiSn and Co$_{2-x}$Ni$_x$TiSn[32]. The TE devices with high reliability are usually found to be cost-prohibitive. Efficiency is usually found to be low. In recent times, higher efficiency has been achieved in terms of high figure of merit being around ZT=6 in doped Fe$_2$V$_{0.8}$W$_{0.2}$Al[33]. Full Heusler alloys like Co$_2$MnSi and Co$_2$MnSn were found to be promising with operating temperatures in the range 400 K-900 K[34–36]. Here we explore CoVZrAl, CoFeZrAl and CoIrZrAl in search of new Heusler alloys with good thermoelectric properties. We investigate the electronic structures of CoVZrAl, CoFeZrAl and CoIrZrAl and explore their formation, dynamical stability, crystal structure and magnetic properties using the first-principles technique. The corresponding TE properties (Seebeck coefficient, electrical conductivity, thermal conductivity, power factor) are also studied by using semi-classical Boltzmann transport theory under relaxation time approximation. Feasibility of synthesis of CoVZrAl, CoFeZrAl and CoIrZrAl and their thermo-

cream
black

[a] Department of Physics, West Bengal State University, Barasat, Kolkata-700126, India; E-mail: molly@wbsu.ac.in
[b] Department of Physics, Saheed Nural Islam Mahavidyalaya, North 24 Pgs., West Bengal-743286, India

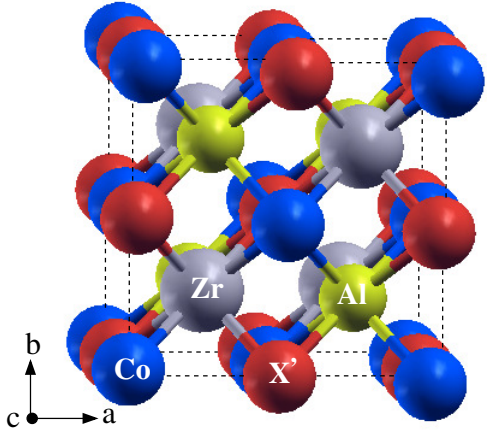

**Fig. 1** (colour online) Crystal structure of quaternary Heusler alloys CoX'ZrAl (X'=V, Fe, Ir).
cream

electric potential are worth exploring.

## 2 Computational details

### 2.1 Electronic structure calculation

We have performed the first-principles electronic structure calculations in the framework of density functional theory (DFT) by using projected agumented wave (PAW) pseudopotentials basis which is implemented in the plane-wave based Vienna ab initio simulation package (VASP) program[37–40]. Generalized gradient approximation (GGA) as implemented in the Perdew-Burke-Ernzerhof (PBE) formalism is used as exchange correlation functional[41]. To compute the structural stability and electronic properties, a $24 \times 24 \times 24$ Monkhorst-pack k-mesh is used for the Brillouin zone (BZ). The cutoff energy for the plane waves is set at 267.968 eV. We have used the condition that the self-consistent (SCF) cycles will converge when the energy difference between two energy cycles is less than $10^{-4}$ eV.

### 2.2 Boltzmann transport theory: Calculation of thermoelectric properties

The thermoelectric properties mainly described by electrical conductivity ($\sigma/\tau$), electronic thermal conductivity ($\kappa_e/\tau$) and Seebeck coefficient (S) are determined using semi-classical Boltzmann transport theory (BTT) under relaxation time ($\tau$) approximation. BoltzTraP2[42] is a program dedicated for interpolating band structures and calculating semi-classical transport coefficients using BTT. Transport coefficients are evaluated from the group velocity of the band (Bloch) electrons. The group velocity corresponding to the $n^{th}$ energy band $\varepsilon_{n,\vec{k}}$ and the $\alpha^{th}$ component of the wave vector $\vec{k}$ is given by,

$$v_\alpha(n,\vec{k}) = \frac{1}{\hbar}\frac{\partial \varepsilon_{n,\vec{k}}}{\partial k_\alpha} \qquad (1)$$

The electrical conductivity tensor, electronic thermal conductivity and Seebeck coefficient tensor are respectively,

$$\sigma_{\alpha\beta}(T,\mu) = \frac{e^2}{N\Omega}\sum_{n,\vec{k}}\int v_\alpha(n,\vec{k})v_\beta(n,\vec{k})\delta(\varepsilon - \varepsilon_{n,\vec{k}})$$

$$\times \left[-\frac{\partial f(T,\mu,\varepsilon)}{\partial \varepsilon}\right]\tau(\varepsilon(\vec{k}))d\varepsilon, \qquad (2)$$

$$(\kappa_e)_{\alpha\beta}(T,\mu) = \frac{1}{N\Omega T}\sum_{n,\vec{k}}\int v_\alpha(n,\vec{k})v_\beta(n,\vec{k})(\varepsilon - \mu)^2$$

$$\times \delta(\varepsilon - \varepsilon_{n,\vec{k}})\left[-\frac{\partial f(T,\mu,\varepsilon)}{\partial \varepsilon}\right]\tau(\varepsilon(\vec{k}))d\varepsilon \qquad (3)$$

and

$$S_{\alpha\beta}(T,\mu) = \frac{e}{N\Omega T \sigma_{\alpha\beta}(T,\mu)}\sum_{n,\vec{k}}\int v_\alpha(n,\vec{k})v_\beta(n,\vec{k})(\varepsilon - \mu)$$

$$\times \delta(\varepsilon - \varepsilon_{n,\vec{k}})\left[-\frac{\partial f(T,\mu,\varepsilon)}{\partial \varepsilon}\right]\tau(\varepsilon(\vec{k}))d\varepsilon \qquad . \qquad (4)$$

where $\alpha, \beta$ are the tensor indices for the thermoelectric parameters and $\Omega$, $\mu$, $f$ and N are the volume of unit cell, chemical potential, Fermi-Dirac distribution function and the number of k points sampled respectively. For the calculation of thermoelectric properties within a constant relaxation time approach, the required input parameters $v_\alpha$, $v_\beta$, $\varepsilon$, $\mu$ are obtained from the output of corresponding DFT calculation. We have computed the thermoelectric properties of these three compounds within a large temperature range of 200-900 K.

### 2.3 Electrical relaxation time calculations

To calculate the electron relaxation time, we used the following expression for $\tau$, derived from the Heisenberg uncertainty principle and the Drude model[43]. For metals and semiconductor $\tau$ can be written as

$$\tau = \frac{eSh}{k_B^2 T} \qquad (5)$$

and

$$\tau = 0.1\frac{eSh}{k_B^2 T} \qquad (6)$$

where e, S are the electronic charge and Seebeck coefficient and h, $k_B$ and T are the Planck's constant, Boltzmann constant and absolute temperature respectively.

### 2.4 Phonon calculations

Phonon calculations are performed using the finite displacement method (FDM) as implemented in VASP, employing a $2 \times 2 \times 2$ supercell and a $3 \times 3 \times 3$ k-point mesh for sampling the BZ of the cubic unit cell. The dynamical matrices are evaluated using the phonopy[44,45] code to obtain the phonon dispersion and phonon density of states. To compute the dynamical matrices, phonopy performs a Fourier interpolation of real-space force constants obtained from the DFT calculations. The phonon density of states (DOS) and phonon band structure of CoX′ZrAl (X′=V, Fe, Ir) are plotted with respect to phonon frequencies and high-symmetry k-points of the BZ, respectively.

### 2.5 Phonon Boltzmann transport equations

To calculate the lattice thermal conductivity tensor ($\kappa_l$), one requires third-order anharmonic interatomic force constants along with second-order harmonic interatomic force constants. The second-order interatomic force constants are computed from the phonon dispersion using phonopy code. For the calculation of third-order interatomic force constants, we have used phono3py [44,46] code using the same supercell size and k-point mesh as that of second-order interatomic force constants calculations. Phono3py generate 452 number of supercell with different atomic displacements for each compounds. The force on each supercell is obtained from DFT calculation, which takes very long time. The $2 \times 2 \times 2$ supercell is sufficient to calculate $\kappa_l$. The $\kappa_l$ is obtained by solving the phonon Boltzmann transport equation (BTE) using the relaxation time approximation as implemented [47] in the phono3py code.

## 3 Results and discussions

### 3.1 Structural parameters and stability analysis

The quaternary Heusler alloys CoVZrAl, CoFeZrAl and CoIrZrAl are obtained by 50% doping at the Co site of $Co_2ZrAl$. Full Heusler compounds with composition $X_2YZ$, are considered to crystallize in the cubic $L2_1$ structure with space group Fm-3m (space group no. 225) as that of $Co_2ZrAl$ [11]. The $L2_1$ structure consists of four inter-penetrating face-centered cubic (fcc) sublattices of the two-X, Y and Z atoms respectively. The quaternary Heusler alloy (XX′YZ) crystallizes in a fcc structure with space group F-43m (216). This type of Heusler compounds have been usually found in the most stable configuration when X, Y, X′ and Z atoms are at Wyckoff positions (0, 0, 0), (1/4, 1/4, 1/4), (1/2, 1/2, 1/2) and (3/4, 3/4, 3/4) respectively. The crystal structure of quaternary Heusler alloys CoX′ZrAl (X′=V, Fe, Ir) is shown in Fig. 1. We have taken the basis for all three from their parent material $Co_2ZrAl$ in the ambient condition and replaced one of the Co atom at (1/2, 1/2, 1/2) site with X′ (where X′=V, Fe, Ir) atom. Subsequently we have varied the lattice parameters in each case, keeping the corresponding basis as unchanged and calculated its total energy. This led to the equilibrium lattice constant (a in Å), bulk modulus in equilibrium condition ($B_0$ in GPa) and $1^{st}$ derivative of bulk modulus with respect to pressure ($B_0'$) by fitting these total energy-volume data (calculated) with Birch-Murnaghan equation of state (EOS) [48] given as,

$$E(V) = E_0 + \frac{9V_0 B_0}{16} \times \left[ \left( \frac{V_0}{V} \right)^{2/3} - 1 \right]^3 B_0' +$$

$$\frac{9V_0 B_0}{16} \left[ \left( \frac{V_0}{V} \right)^{2/3} - 1 \right]^2 \left[ 6 - 4 \left( \frac{V_0}{V} \right)^{2/3} \right] \quad (7)$$

where V is the volume of unit cell and $V_0$ is the volume at equilibrium condition. The total energy-volume plots (dots for calculated and line joining the fitted values) for these compounds are shown in Fig. 2. The calculated values of 'a', $B_0$, $B_0'$ have been listed in Table 1 . We observe that the lattice constant of CoFeZrAl is considerably smaller than that of CoVZrAl and CoIrZrAl at equilibrium. Bulk modulus of a material is the measure of the ability of a material to be stable when its volume changes under pressure. The bulk modulus at equilibrium increases with increase in atomic number of X′ (X′= V, Fe, Ir) atom. The positive values of $B_0'$ imply that these materials are rigid under pressure. The calculated pressure derivative of bulk modulus ($B_0'$) of these compounds are almost similar.

Total energies are calculated for CoX′ZrAl in their equilibrium state as well as for the individual components separately. The formation energy is given by,

$$\Delta E_0 = E_{tot} - (E_X - E_{X'} - E_Y - E_Z) \quad (8)$$

where $E_{tot}$ is the total energy of CoX′ZrAl (X′=V, Fe, Ir) and $E_X$, $E_{X'}$, $E_Y$, $E_Z$ are the total energy of individual Co, X′, Zr and Al solids respectively. The calculated formation energy of CoIrZrAl, CoFeZrAl and CoVZrAl are shown in Table 1 [49]. The negative formation energy of all three compounds indicates to their thermodynamic stability, suggesting that their formation is energetically favorable.

### 3.2 Dynamical stability

Phonons play a crucial role in determining the dynamical stability of crystal structures. The structural stability, along with the vibrational contribution to the thermal properties of alloys, can be analyzed using phonon dispersion spectra. Fig. 3 presents the phonon dispersion curves plotted along the high-symmetry path Γ → L → W → X→ Γ in the BZ of the crystal structure and phonon density of states for CoX′ZrAl (X′ = V, Fe, Ir). The phonon dispersion relations are obtained using the finite displacement method, which uses the forces acting on atoms in a supercell after individually displacing each atom. Phonopy explicitly evaluates the dynamical matrices from the forces computed via the Hellmann-Feynman theorem as implemented in the DFT calculations in VASP. No phonon anomalies are observed in the phonon DOS for any of the three compounds. The presence of real and positive phonon frequencies confirms the dynamical stability of the structures. The phonon dispersion curve consists of 12 branches, comprising 3 acoustic and 9 optical branches. The three lowest-energy phonon modes correspond to the acoustic modes and are classified into TA and LA. The TA and LA modes correspond to two transverse and one longitudinal vibrations respectively, both displaying linear dispersion near the Γ point. From the partial phonon density of states (DOS), we observe that heavier atoms contribute predominantly in the low-frequency region while lighter atoms play a dominant role at higher frequencies. For both CoVZrAl and CoFeZrAl compounds, the Zr atom contributes the most to the acoustic modes, whereas in CoIrZrAl, the Ir atom plays a major role. In the low-frequency optical modes, the Co atom has the highest contribution in both CoVZrAl and CoFeZrAl. However, in CoIrZrAl, both Co and Ir atoms contribute significantly. For all three compounds, the Al atom has the highest contribution to the high-frequency optical modes. Additionally, we observe that the high-frequency branches are narrower compared to the low-frequency branches.

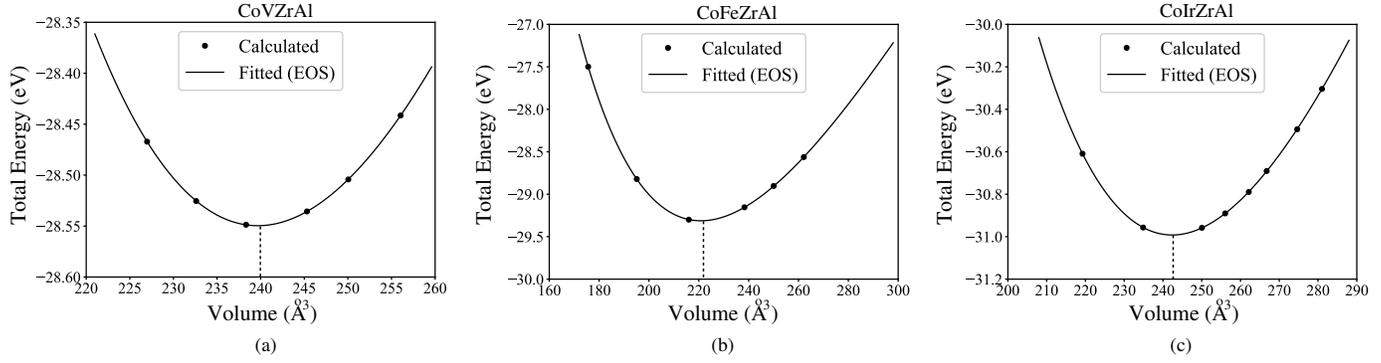

Fig. 2 Total energy-volume curve for unit cell of (a) CoVZrAl (b) CoFeZrAl (c) CoIrZrAl. The symbols denote the DFT calculated total energy for different volume of unit cell and the total energy-volume fitting was done using Birch-Murnaghan EOS (7).

Table 1 Calculated structural parameters and formation energy for CoX′ZrAl (X′ = V, Fe, Ir) compounds.

| Compounds | a(Å) | $B_0$(GPa) | $B'_0$ | $\Delta E_0$(eV) |
|---|---|---|---|---|
| CoVZrAl | 6.21 | 35.25 | 4.92 | -0.5683 |
| CoFeZrAl | 6.04 | 42.05 | 4.22 | -2.6091 |
| CoIrZrAl | 6.23 | 46.48 | 4.33 | -3.1066 |
| cream | | | | |

### 3.3 Electronic structure and magnetic properties

The feasibility of synthesizing these materials is tested by calculating their formation energy in their equilibrium state. The equilibrium crystal structure is determined, the details of which is given in section 3.1. Also to be noted is the fact that CoVZrAl has already been predicted[31] to be a magnetic semiconductor but is yet to be synthesized till date. All the three compounds possess a crystal structure of cubic symmetry, but their lattice parameters differ, as shown in the Table 1.

Before discussing their electronic structures, we shed light on the magnetic ground state of each material. The magnetic ground state is found by comparing total energies of each material in its non-magnetic, ferromagnetic and an antiferromagnetic (AFM) spin configuration (AFM1) with a $2 \times 2 \times 2$ supercell. The AFM1 spin configuration is the conventional one with opposite spin on nearest neighbour ions. The compound CoFeZrAl turns out to be non-magnetic (NM). The lowest energy state for the other two is a ferromagnetic one. The energy difference $\Delta E$ ($E_{FM}$-$E_{AFM}$) between FM ground state and AFM1 state of CoVZrAl and CoIrZrAl is given in Table 2. A crude estimation of Curie temperature has been calculated and shown in Table 2 using mean-field approximation (MFA). We have used the formula, $T_c = \frac{2}{3k_B}\Delta E$ to calculate the phase transition temperature of the magnetic compounds. The calculated $T_c$ of CoVZrAl close to the early reported value of 892 K using MFA and 652 K using RPA[31].

Hence the self-consistent electronic structures in the FM phase are calculated and discussed for all three alloys. The spin-polarized band structures of CoX′ZrAl (X′ = V, Fe, Ir) in their ferromagnetic phase along high symmetry lines in the reciprocal space for both spin-up and spin-down channels have been shown in Fig. 4(a), Fig. 4(b) and Fig. 4(c) respectively. The accuracy and reliability of the thermoelectric properties depends on the accuracy of the electronic band structures. Hence a fine Brillouin zone has been choson and optimized. We get the necessary information on the electronic nature of the materials, as to whether the material is a metal or a semiconductor or an insulator and also the value of the band gap if any, from the band structure diagram. Fig. 4(a) shows that the compound CoVZrAl has a band gap of about 0.17 eV in the spin up channel and 1.08 eV in spin-down direction and is therefore a direct gapped material. The material CoFeZrAl shows no band gap in both spin channels as per Fig. 4(b). The compound CoIrZrAl is metallic in the spin-up channel whereas the spin-down channel has a small band gap about 0.34 eV as shown in Fig. 4(c). Apparently the band structure of CoVZrAl and CoFeZrAl are similar in many accounts. The lowest occupied valence band around -6 eV is derived from Al-3s electrons. This is followed by Al-3p electrons hybridized with the Co-3d and V-3d/Fe-3d electrons higher up in energy. Let us first try to analyze the crystal field splitting pattern in these systems. For CoVZrAl compound, the ligands around Co are two Zr and two Al ions forming square planes above and below the Co ion respectively, as if mimicking a cubic crystal electric field (CEF) but with 4 direct Co-Al interactions and 4 less direct Co-Zr interactions. Hence we expect a crystal field splitting midway between cubic and octahedral, leading to Co-3d state splitting into lower $e_g$ and higher $t_{2g}$ for Co and same for V/Fe/Ir. However Zr-4d electrons behave differently. These Zr ions have a cubic coordination formed by an alternate arrangement of Co and V/Fe/Ir ions on the corners of the hypothetical cube around it. Hence the localized 4d electrons of Zr atom are surrounded by more localized 3d electrons of Co and V atoms in CoVZrAl. A direct d-d interaction is weak here. The Zr-4d states will thus split under a

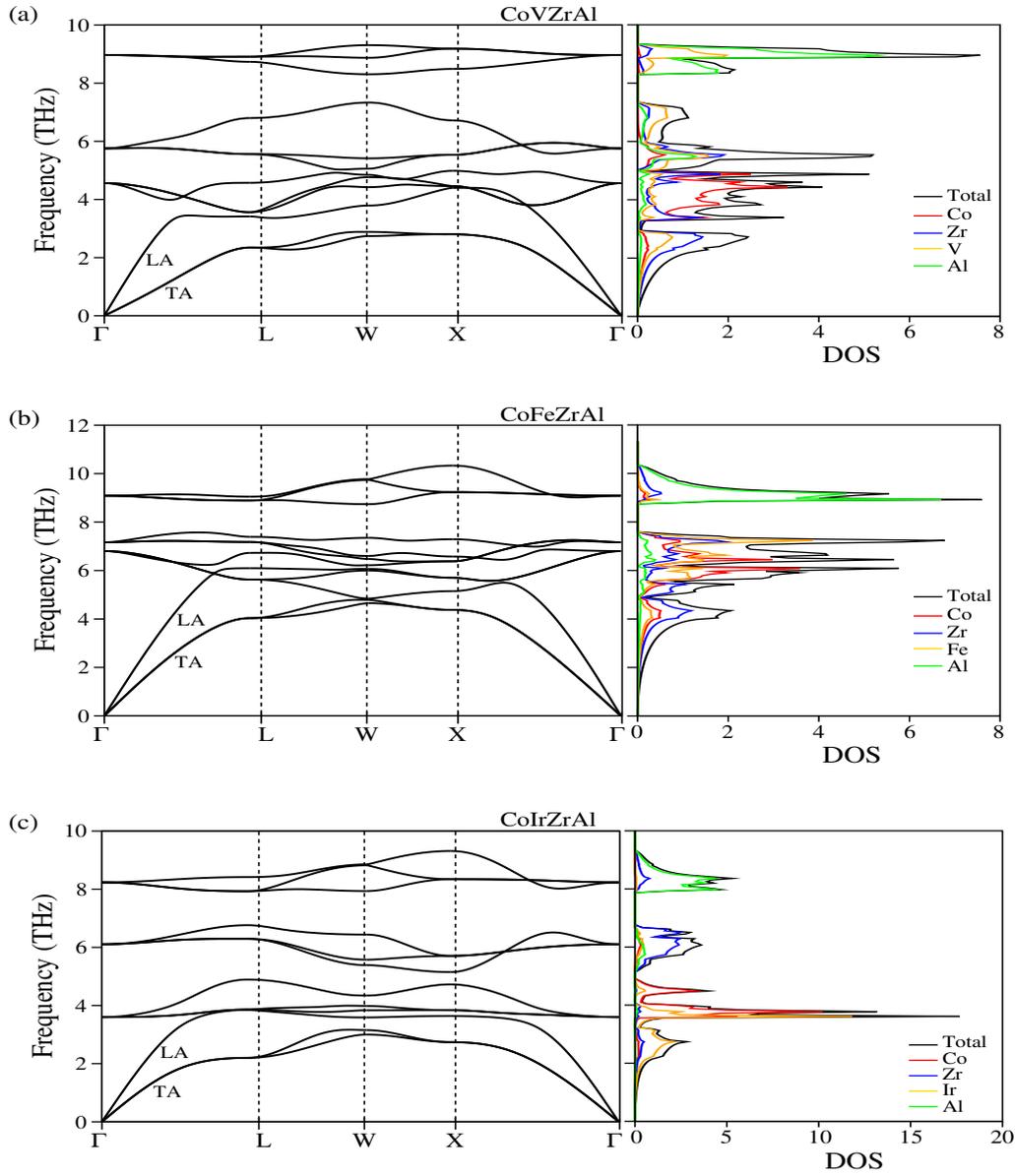

Fig. 3 (colour online) The phonon dispersions (left) and phonon density of states (right) for (a) CoVZrAl (b) CoFeZrAl (c) CoIrZrAl. In phonon dispersion TA, LA represents the acoustic branches.
cream

weak cubic or tetrahedral field into lower doubly degenerate $e_g$ and higher triply degenerate $t_{2g}$ bands. The hybridized Co-X' 'd' states will generate two lower bonding states i.e. $e_g$ and $t_{2g}$ and two higher anti-bonding states i.e. $e_g^*$ and $t_{2g}^*$. The 'd' electrons of Zr atom hybridize with these hybrid states of Co-X' atoms yielding two lower bonding states $e_g$ ((Co-X')-Zr) and $t_{2g}$ ((Co-X')-Zr) and two anti-bonding states $e_g^*$ ((Co-X')-Zr) and $t_{2g}^*$ ((Co-X')-Zr) and two non-bonding states (Co-X')-$e_u$ (doubly degenerate) and (Co-X')-$t_{1u}$ (triply degenerate). The $e_g$, $t_{2g}$, $t_{1u}$ and $e_u$ states are indicated as p, q, r and s respectively in band structure of CoIrZrAl [Fig. 4(c)] only for the $\Gamma$ point in the minority spin channel.

The spin-polarized total density of states (TDOS) and partial density of states (PDOS) for both the spins for all three are presented in Fig. 5. It is clear from the TDOS in Fig. 5(a) that CoV-ZrAl is a much-desired semiconductor in the ferromagnetic state.

Hence it has high potential for an ideal spin filter. The compound CoFeZrAl is a non-magnetic metal and CoIrZrAl behaves like a half-metal according to Fig. 5(b) and Fig. 5(c). We can explain TDOS by dividing it into three distinct parts — two valence band manifolds (lower and upper) and a conduction band in both spin channels. We observe that, for all the three compounds, the lower and upper valence band regions lie from -5 eV to -1 eV and -1 eV to $E_F$ respectively in the two spin channels. The gap in the minority spin channel closes in CoFeZrAl and the DOS in the minority spin channel is distinctly different from that in the other two. The crystal-field split hybridized Co-3d states occur at around -2.5 eV and the V-3d/Fe-3d states occur at -1 eV. These two states are reversed in hierarchy as X' is replaced by Ir. Both these states hybridize strongly with Zr-4d states to form the bonding manifold states. In between these states, the Al-3p state spreads out. The non-bonding (Co-3d-V/Fe-3d) $t_{1u}$ state lies slightly below $E_F$ in up spin channel of CoVZrAl and the same is unoccupied in the down spin channel. This leads to gaps in both spin channels of around 0.17 eV (in up spin channel) and 1.08 eV (in down spin channel) respectively. Interestingly CoFeZrAl is not only non-magnetic but the $E_F$ lies in a pseudogap formed by $t_{1u}$ and $e_u$ (non-bonding states of Co-3d/Fe-3d) states. The Co-3d manifold is spread from -3 eV to $E_F$ whereas the Ir-5d states from around -6 eV with peak at about -3.5 eV to $E_F$ in spin up channel in CoIrZrAl. This is on account of the fact that Ir is a $5d^7$ system and has delocalized electrons whereas V and Fe are $3d^3$ and $3d^6$ which are more localized than 5d states. This implies that larger hybridization in CoIrZrAl which we observe in DOS [Fig. 5(c)] of CoIrZrAl.

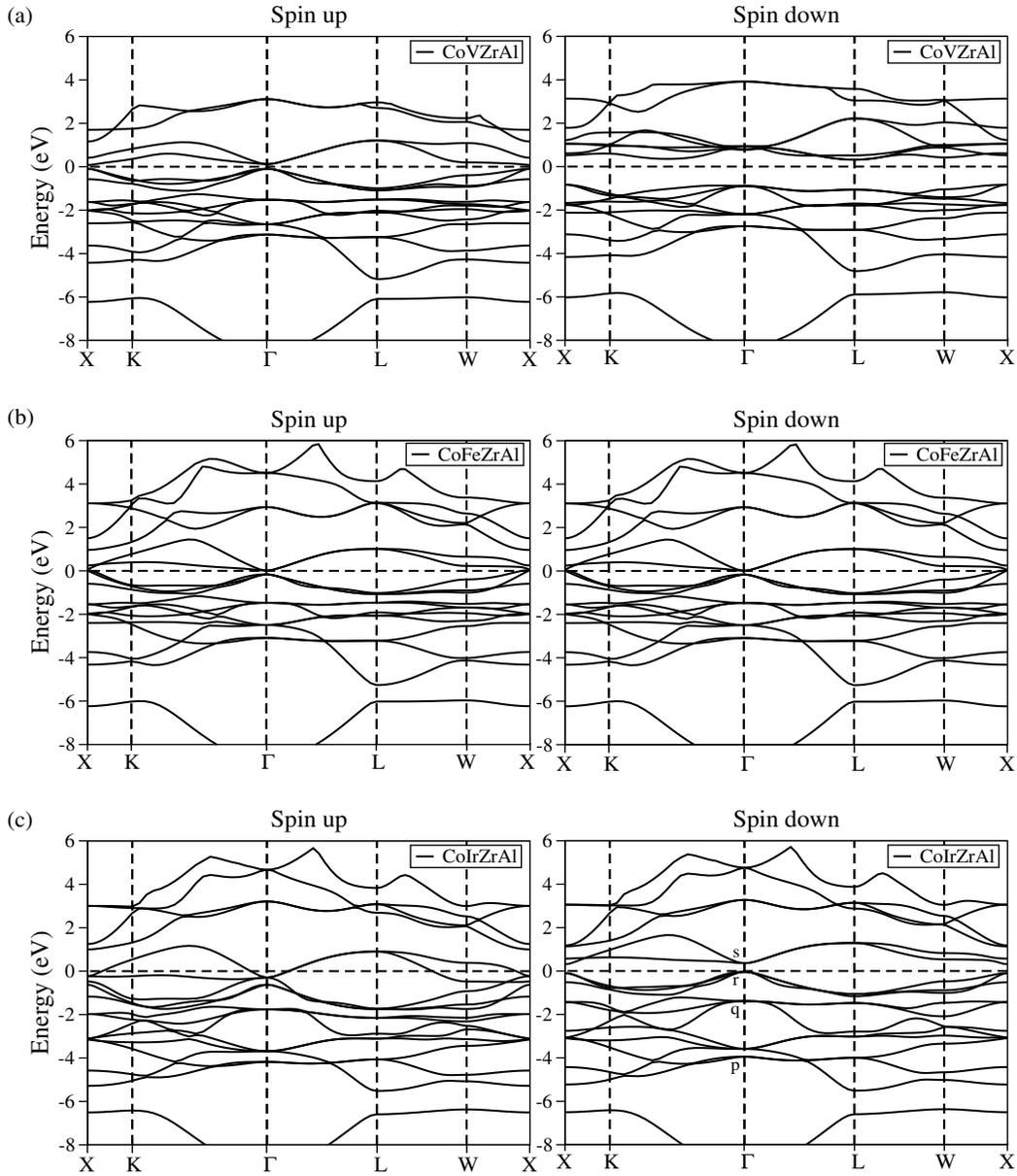

Fig. 4 Calculated spin-up and spin-down band structures of FM (a) CoVZrAl (b) CoFeZrAl (c) CoIrZrAl.
cream

Let us try to analyze and relate the varying electronic and magnetic properties with varying $X'$ element. The non-bonding $e_u$ state derived from Co-3d-V/Ir-5d shows a gradual movement from above $E_F$ to $E_F$ as $X'$ changes from V to Ir in up spin channel whereas it is the movement of the $t_{1u}$ non-bonding state from above $E_F$ in CoVZrAl to below $E_F$ in both CoFeZrAl and CoIrZrAl in the down spin channel that causes the overall change in the electronic structure. The schematic diagram of the band structure of CoIrZrAl (25 valence electrons) at $\Gamma$ point for both channels has been shown in Fig. 6 based on the above discussion. The material CoIrZrAl shows half-metallic character because of its small energy gap between $t_{1u}$ and $e_u$ states in spin-down channel whereas the spin-up channel shows a singly filled $e_u$ state. For CoFeZrAl, with 24 valence elecrons, the Fermi level lies between $t_{1u}$ and $e_u$ states for both spin channels. It shows a pseudo-gap like behaviour because of small energy difference and mixing between $t_{1u}$ and $e_u$ states. The $E_F$ in CoVZrAl (with 21 valence electrons) lies between $t_{1u}$ and $e_u$ states for spin-up and between $t_{2g}$ and $t_{1u}$ states for spin-down channel. This will render CoVZrAl semiconduct-

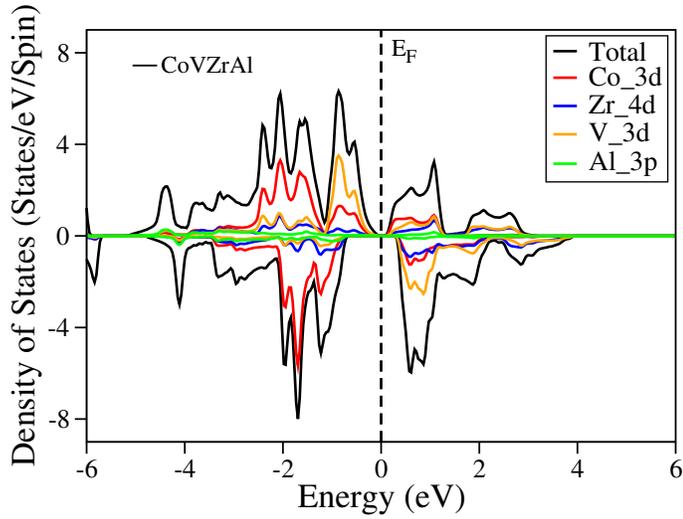

(a)

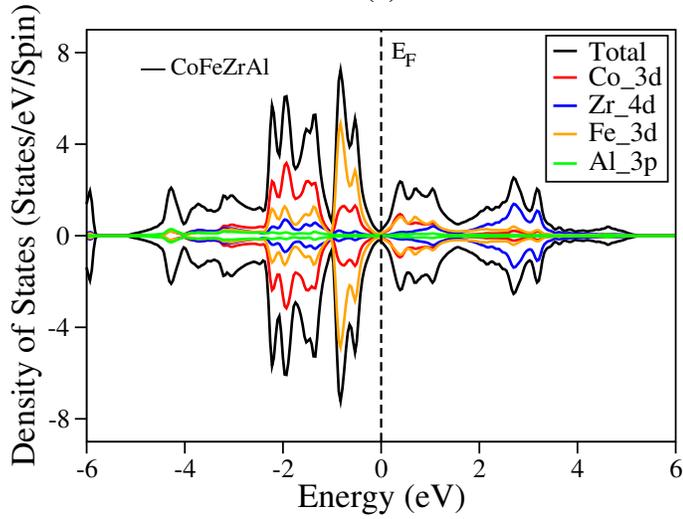

(b)

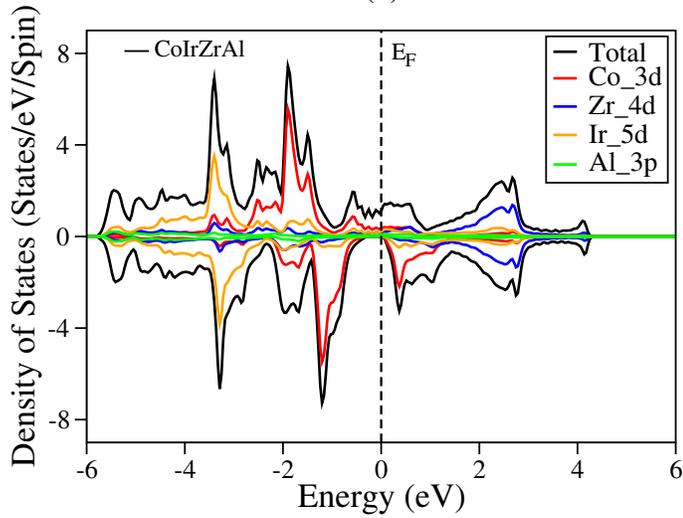

(c)

Fig. 5 (colour online) TDOS and PDOS plots of (a) CoVZrAl (b) CoFeZrAl (c) CoIrZrAl.

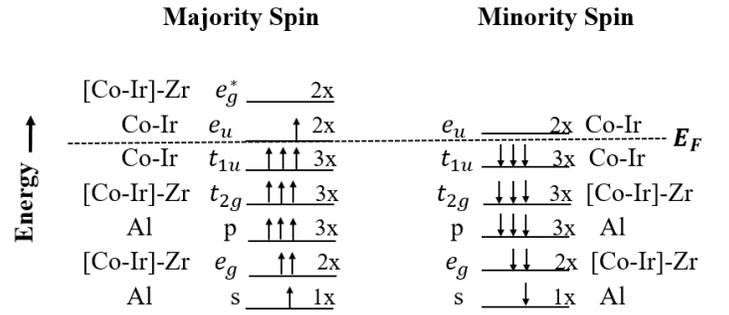

Fig. 6 Schematic diagram of the character of bands of CoIrZrAl at the Γ point for spin-up (majority-spin) and spin-down (minority-spin) electrons. The number factor on the lines represents the degree of degeneracy of the band at the Γ point. The arrow symbol on the lines represents the number of electrons in that energy levels.

ing in nature if it is ultimately synthesized. We observe that the above band structure splitting scheme for CoX'ZrAl is in complete agreement with their calculated band structures.

Within the above-discussed scheme, there are 12 and 9 valence electrons in the spin-up and spin-down channels respectively for CoVZrAl. So the total moment should be close to 3 $\mu_B$ according to Slater-Pauling rule. We also observe that the fully filled non-bonding t$_{1u}$ state of majority spin channel derived from both Co-3d and V-3d states and has no substantial contribution from either Zr or Al atoms. The total and atom-decomposed magnetic moments are given in Table 2. The less than 3 $\mu_B$ total moment calculated is due to hybridization-induced loss of electrons from the spin-up channel. Similarly, there are 12 valence electrons in each spin channel in CoFeZrAl. This leads to a non-magnetic ground state despite the presence of the Fe atom. The small density of states at E$_F$ is again hybridization-induced. The deep pseudo-gap at E$_F$ is very similar to that found in graphene. This is a indicator of CoFeZrAl to have high electrical conductivity. Finally CoIrZrAl has 13 and 12 valence electrons in the two spin channels. This leads to a total magnetic moment of 1 $\mu_B$ which arises from the Co-3d and Ir-5d driven half filled e$_u$ state in the spin-up channel. Hence we find perfect harmony between the suggested splitting scheme and calculation. The major source of magnetism in CoV-ZrAl are the localized 3d electrons of Vanadium whereas Ir atoms do not contribute to the local moment but Co does in CoIrZrAl. However magnetism is a combined effect of local moments and their action between these local moments. From DOS [Fig. 5] we observe the interaction between local moments of Co-V/Co-Ir happens via hybridization with Zr-4d electrons. The CoVZrAl has large moment which is much smaller in CoIrZrAl and the hybridization in CoVZrAl is smaller than in CoIrZrAl. This explain nearly similar T$_c$ for both.

### 3.4 Thermoelectric properties

The energy conversion efficiency of TE materials at a given temperature T is estimated by a dimensionless quantity called thermoelectric figure of merit (ZT). The ZT of a TE material at a given temperature T is given by $ZT = S^2\sigma T/\kappa$, where S, $\sigma$, $\kappa = \kappa_e + \kappa_l$ are the Seebeck coefficient, the electrical conductivity and to-

Table 2 Calculated total, atom decomposed spin moments (in $\mu_B$), energy difference $\Delta$E (in eV) between two lowest energy states and mean-field estimation of Curie temperatures $T_c$ (in K) for CoX'ZrAl (X' = V, Fe, Ir) compounds.

| Compounds | $m_{Co}$ | $m_{X'}$ | $m_{Zr}$ | $m_{Al}$ | $m_{tot}$ | $\Delta$E | $T_c$ |
|---|---|---|---|---|---|---|---|
| CoVZrAl | 0.184 | 2.247 | 0.241 | 0.014 | 2.69 | 0.089 | 688 |
| CoFeZrAl | -0.000 | -0.001 | 0.001 | 0.000 | -0.00 | - | - |
| CoIrZrAl | 0.978 | 0.150 | -0.069 | -0.014 | 1.05 | 0.087 | 673 |

cream

tal thermal conductivity that consists of both electronic ($\kappa_e$) and phonon ($\kappa_l$) contribution to the thermal conductivity. The electrical conductivity is given by $\sigma = ne\mu$, where e is the electronic charge, n and $\mu$ are the number density and mobility of charge carriers respectively. In order to achieve high electrical conductivity, the mobility of charge carriers must be high. The mobility of electrons and holes are given by $\mu_e = \frac{e\tau}{m_e^*}$ and $\mu_h = \frac{e\tau}{m_h^*}$ respectively, where $\tau$ is the relaxation time, $m_e^*$ and $m_h^*$ are the effective masses of electrons and holes. A higher effective mass implies a lower mobility of the charge carriers and hence lower electrical conductivity. A dispersionless (or localized) band structure gives higher effective mass and the more dispersive band structure yields smaller effective mass. One encounters a combination of both light and heavy charge carriers in a system. Light charge carriers increase the mobility as well as electrical conductivity but the heavy charge carriers assume the major role of enhancing the Seebeck coefficient [Eq. 4]. A larger value of Seebeck coefficient also gives a higher value of power factor which results in a higher thermoelectric figure of merit.

BoltzTraP2 computes the electronic transport properties, mainly the Seebeck coefficient (S), electrical conductivity for constant relaxation time ($\sigma/\tau$), thermal conductivity due to electrons for constant relaxation time ($\kappa_e$), power factor (P.F) for constant relaxation time ($S^2\sigma/\tau$) and figure of merit ($ZT = S^2\sigma T/\kappa$) but it assumes that the phonon contribution to lattice thermal conductivity ($\kappa_l$) is zero. The electronic relaxation time and lattice thermal conductivity need to be estimated. The calculated electronic relaxation time using Eq. 5 and Eq. 6 is shown in Fig. 7. We have used the calculated equilibrium lattice constants for the calculation of TE properties. We have computed the thermoelectric parameters for CoX'ZrAl (X' = Fe, V, Ir) from 200 K to 900 K temperature range and plotted them with temperature.

Seebeck coefficient gives the information about the nature of dominant charge carriers present in the material. For semiconductor, a positive Seebeck coefficient implies that holes are the dominating carriers while its negative value represents the dominance of electrons. The variation of Seebeck coefficient with temperature for all the three compounds aforementioned has been shown in Fig. 8(a). Among these three compounds, CoVZrAl has the highest value of Seebeck coefficient $\sim -538\mu V/K$ at 200 K in n-region. Hence dominant carriers are electrons. This can be clearly observed in the band structure diagram [Fig. 4(a)]. At about 200 K, the band gap closes, thereby allowing electrons to move into the conduction band, where they become the dominant charge carriers. The highest value of Seebeck coefficient for CoFeZrAl is $\sim -112\mu V/K$ at 500 K and $\sim -19.2\mu V/K$ at 900 K for CoIrZrAl. We observe that the Seebeck coefficient (S) of CoVZrAl gradually decreases with increasing temperature, whereas for CoIrZrAl, it increases with temperature and remains nearly constant at higher temperature. In the case of CoFeZrAl, the Seebeck coefficient initially increases and then decreases with rising temperatures.

Fig. 8(b) shows the variation of electrical conductivity ($\sigma$) with temperature for all three compounds. At 200 K, CoFeZrAl has a highest $\sigma$ value of 5.83×10$^6$ $\Omega^{-1}$m$^{-1}$ in comparison with the other two compounds, as already anticipated while discussing the DOS at $E_F$ as seen in Fig. 5(b). Electrical conductivity decreases with rising temperatures indicating metallic behavior. The temperature-dependent electronic part of thermal conductivity ($\kappa_e$) and lattice thermal conductivity ($\kappa_l$) have been presented in Fig. 9. In order to obtain the maximum thermoelectric efficiency, $\kappa$ must be minimum. As expected, both $\kappa_e$ and $\kappa_l$ are minimum for CoVZrAl because of its semiconducting nature as compared to the other two compounds. We also observe that for all three compounds, the $\kappa_e$ increases with temperature, indicating a metallic character. The lattice thermal conductivity ($\kappa_l$) decreases with temperature for all the three compounds. Maximum Seebeck coefficient and minimum $\kappa$ [Fig. 10(a)] occur for CoVZrAl which also has an optimum electrical conductivity. Hence we find that the desired thermoelectric coefficients leading to maximum figure of merit are obtained for CoVZrAl. The variation of P.F. ($S^2\sigma$) for CoX'ZrAl (X'=V, Fe, Ir) with temperatures is presented in Fig. 10(b). The compound CoFeZrAl has the highest P.F. at 900 K because of its corresponding high electrical conductivity ($\sigma$) [Fig. 8(b)]. We also find that the P.F. increases with temperature for all the three compounds. We can also analyze the P.F. from band structure. From Fig. 4, we observe degener-

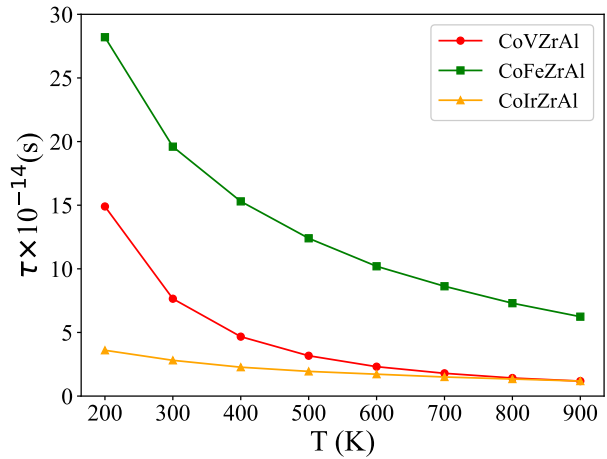

Fig. 7 (color online) The temperature-dependent electronic relaxation time of CoVZrAl, CoFeZrAl and CoIrZrAl compounds.

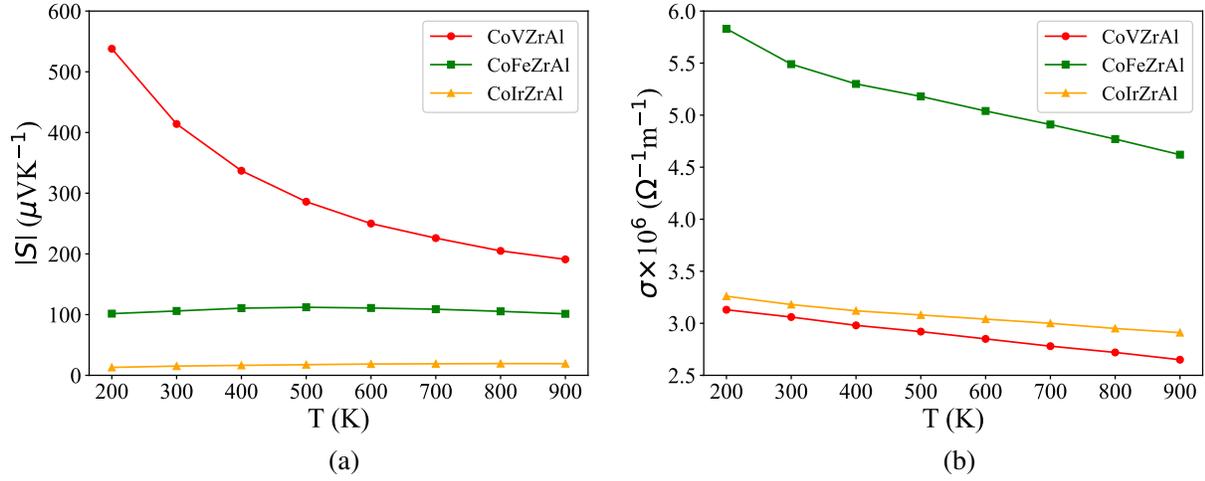

Fig. 8 (color online) The temperature-dependent (a) Seebeck coefficient (S) and (b) electrical conductivity ($\sigma$) of CoVZrAl, CoFeZrAl and CoIrZrAl compounds.

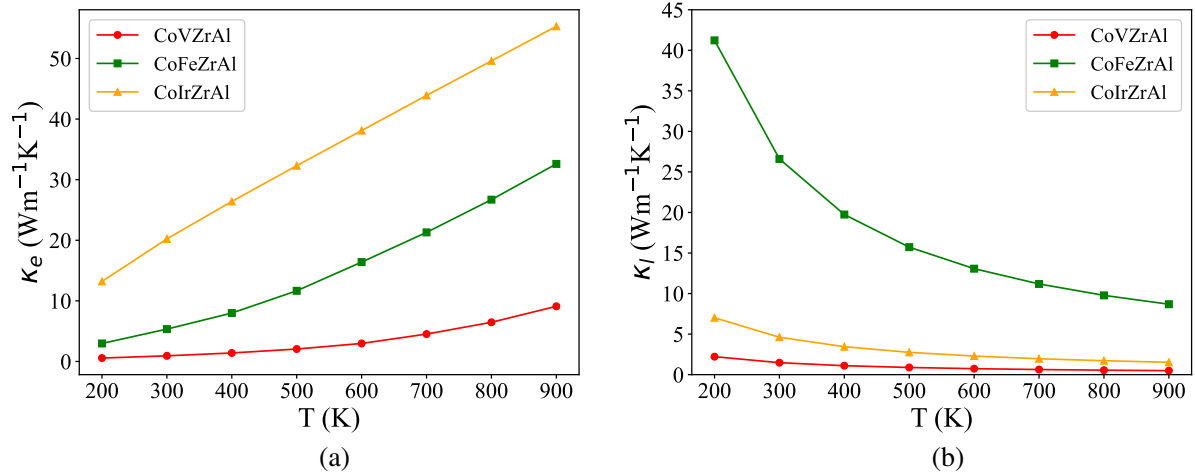

Fig. 9 (color online) The temperature-dependent (a) electronic thermal conductivity ($\kappa_e$) and (b) lattice thermal conductivity ($\kappa_l$) of CoVZrAl, CoFeZrAl and CoIrZrAl compounds.
cream

acy of both valence band maximum (VBM) and conduction band minimum (CBM) at $\Gamma$ point. Among the VBM and CBM at $\Gamma$ point, the more dispersive bands provide lighter charge carriers and flat band provides heavier charge carriers. So we get a combination of both light and heavy charge carriers. Light charge carriers contribute to increase in the electrical conductivity of the material whereas heavy charge carriers contribute to enhance the Seebeck coefficient. Hence an optimized combination of these two factors increases the thermoelectric power factor. The P.F. of CoIrZrAl is very small due to its small value of Seebeck coefficient. On the other hand, the P.F. of CoVZrAl, although smaller than that of CoFeZrAl, assumes an optimum value.

Based on the above calculated transport parameters, we have estimated the thermoelectric figure of merit of CoX'ZrAl (X'= V,Fe,Ir) from 200 K to 900 K tempearure range. At high temperature P.F. is large for CoFeZrAl than CoVZrAl but $\kappa$ is smallest for CoVZrAl and largest for CoFeZrAl. This makes ZT largest for CoVZrAl and smallest for CoFeZrAl. The variation of ZT with temperature has been shown in Fig. 10(c). In order to achieve a higher ZT, we require maximum value of power factor ($S^2\sigma$) along with minimum value of thermal conductivity ($\kappa$). Although no theoretical perimeter is set for highest value of ZT, one of the conventions to obtain the higher value of figure of merit by varying the three transport parameters $S, \sigma$ and $\kappa$. For CoFeZrAl and CoIrZrAl, figure of merit increases with increasing temperature because of increasing P.F. but CoVZrAl behaves differently. The ZT of CoVZrAl initially increases with temperature but then it decreases with rising temperature. Therefore among these three compounds, CoVZrAl has the highest ZT value of about 1.4 at 600K. The compound CoFeZrAl has maximum ZT value of 0.31 at 900K and that of CoIrZrAl is very small because of its very low Seebeck coefficient and large electrical thermal conductivity

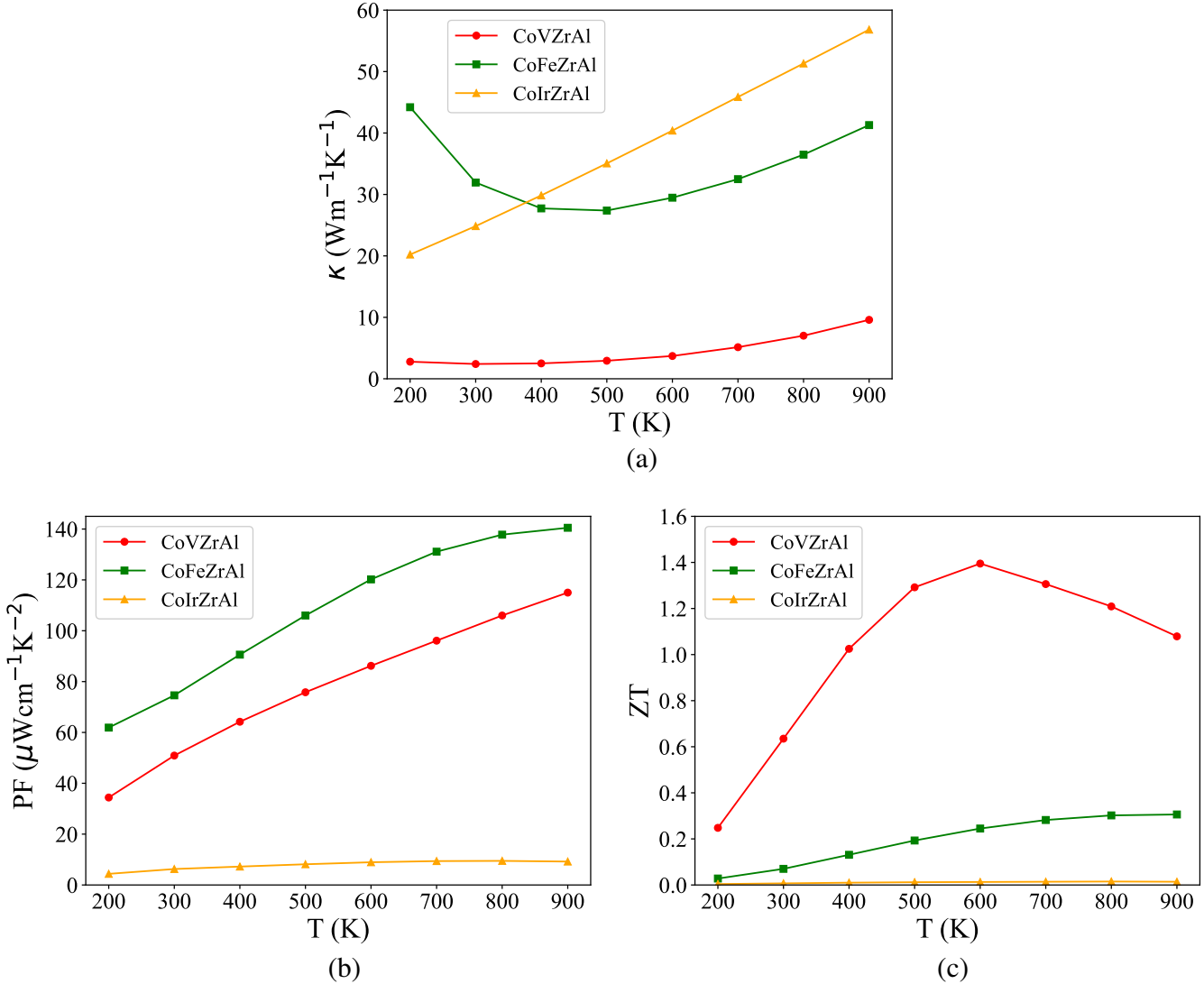

Fig. 10 (color online) The temperature-dependent (a) thermal conductivity ($\kappa$), (b) power factor (PF) and (c) thermoelectric figure of merit (ZT) of CoVZrAl, CoFeZrAl and CoIrZrAl compounds.

cream

($\kappa_e$). If we examine the electronic structures, we can identify the fundamental reason why certain materials possess good thermoelectric properties. The ground state CoVZrAl exhibits a small band gap, resulting in the absence of charge carriers. From Eq. 4, we note that the Seebeck coefficient is inversely related to number of charge carriers, though other factors may also influence this relationship. Theoretically, we expect a large Seebeck coefficient for semiconducting CoVZrAl, which is found to be consistent with our calculations. On the other hand, the band structures of both CoFeZrAl and CoIrZrAl feature a mix of dispersive and flat bands at the Fermi level, indicating the presence of both light and heavy charge carriers. While the band structures of CoFeZrAl and CoIrZrAl have gross similarity, CoFeZrAl exhibits a pseudogap and CoIrZrAl has a small energy gap in the spin-down channel at $E_F$. At 200 K, CoFeZrAl shows a lower Seebeck coefficient than CoVZrAl due to the larger presence of charge carriers in CoFeZrAl at $E_F$ and hence it exhibits a higher electrical conductivity too compared to the other two compounds. At the same temperature, CoIrZrAl has a very small Seebeck coefficient, attributed to the higher number of charge carriers at $E_F$. To achieve the highest figure of merit (ZT), we aim for the lowest possible value of thermal conductivity. However, at high temperatures where the P.F. is large, the lowest value of $\kappa$ is higher for CoIrZrAl than for CoFeZrAl. As a result, the ZT value for CoIrZrAl is significantly lower than that of the other two compounds. This indicates that the presence of delocalized 5d states in transition metals reduces the potential for good thermoelectric performance whereas more localized 3d transition metal enhance the likelihood of a material being a good candidate as a thermoelectric material. Doped semiconductor alloys of Antimony and Bismuth tellurium, having ZT values greater than 1 and operating near room temperature, are the best thermoelectric materials known till date[50]. Previous studies have reported that the magnetic semiconductor CoVTiAl exhibits a ZT of 0.30 at 800 K[51]. In comparison with doped semi-conductor alloys of antimony and bismuth tellurium and also with quaternary Heusler alloy CoVTiAl, CoVZrAl has good potential for applications as a thermoelectric material.

## Conclusion

We have performed the first-principles electronic structure calculation for CoVZrAl, CoFeZrAl and CoIrZrAl. Among these three compounds, CoFeZrAl and CoIrZrAl are studied for the first time. The compound CoVZrAl and CoIrZrAl are ferromagnetic, out of which first shows semiconducting nature and other is a half-metal. The compound CoFeZrAl is non-magnetic but has high electronic mobility. The FM CoVZrAl shows a band gap of 0.17 eV, implying its importance as a spin-filter material and applicability for thermoelectric application because of its large Seebeck coefficient along with small value of thermal conductivity with BTT. The compound CoFeZrAl has a pseudo gap with a large electrical conductivity but also with a large thermal conductivity much like Graphene. The half-metal CoIrZrAl can be useful as a spin-injector. Although CoVZrAl is projected to be a member of rare ferromagnetic semiconductor family, its formation energy suggests that it might be the hardest to synthesize among the three. The compound CoVZrAl has maximum ZT value of about 1.4 at 600 K. We believe that presence of delocalized electrons of transition metals in Heusler alloys leads to degradation of thermoelectric performance of Heusler alloys.

## Conflicts of interest

There are no conflicts of interest to declare.

## acknowledgments

Poulami Biswas would like to acknowledge financial support received from Department of Science & Technology (DST), Govt. of India through its DST INSPIRE Fellowship program vide sanction order no. DST/INSPIRE Fellowship/2021/IF210716. All the three authors acknowledge infrastructural support from Department of Higher Education , Govt. of WB, India. We thank the authors of BoltzTraP2 program for providing access to the code.